\documentclass[aps,pra,floatfix,twocolumn]{revtex4}

\usepackage{amsmath}
\usepackage{epsfig,graphicx}
\usepackage[dvips,ps2pdf]{hyperref}

\newcommand{\bra}[1]{\langle#1|}
\newcommand{\ket}[1]{|#1\rangle}
\newcommand{\be}{\begin{equation}}
\newcommand{\ee}{\end{equation}}
\def\d{\delta}
\def\Om{\Omega}

\begin{document}

\title{Demonstration of a quantum logic gate in a cryogenic surface-electrode ion trap}

\author{Shannon X. Wang}
\email[]{sxwang@mit.edu}
\author{Jaroslaw Labaziewicz}
\author{Yufei Ge}
\author{Ruth Shewmon}
\author{Isaac L. Chuang}
\affiliation{Center for Ultracold Atoms, Department of Physics, Massachusetts Institute of Technology, 77 Massachusetts Avenue, Cambridge, Massachusetts, 02139, USA}

\date{\today}

\begin{abstract}
\noindent We demonstrate quantum control techniques for a single trapped ion in a cryogenic, surface-electrode trap. A narrow optical transition of Sr$^+$ along with the ground and first excited motional states of the harmonic trapping potential form a two-qubit system. The optical qubit transition is susceptible to magnetic field fluctuations, which we stabilize with a simple and compact method using superconducting rings. Decoherence of the motional qubit is suppressed by the cryogenic environment. AC Stark shift correction is accomplished by controlling the laser phase in the pulse sequencer, eliminating the need for an additional laser. Quantum process tomography is implemented on atomic and motional states by use of conditional pulse sequences. With these techniques, we demonstrate a Cirac-Zoller controlled-NOT gate in a single ion with a mean fidelity of 91(1)\%.

\end{abstract}

\maketitle

\section{Introduction}

Trapped ions are promising candidates for realizing large-scale quantum computation \cite{Haeffner:08, Blatt:08}. Significant progress has been made in demonstrating the fundamental ingredients of a quantum processor, with much progress in gate fidelities \cite{Benhelm:08} and multi-ion entanglement \cite{Leibfried:05, Haeffner:05}. In recent years there has been increasing interest in microfabricated surface-electrode traps, owing to their inherent scalability \cite{Seidelin:06, Stick:06}. However, quantum gates have yet to be demonstrated in such systems. An issue with miniaturization of traps is that anomalous heating of the ion's motional state scales unfavorably with trap size \cite{Epstein:07}, potentially limiting gate fidelity in traps of suitable dimensions for scalability \cite{Steane:07}. Recently, it has been shown that by cooling to cryogenic temperatures, the heating rate can be reduced by several orders of magnitude from room-temperature values \cite{Labaziewicz:08}, thus providing one potential solution to this problem. In this work, we demonstrate a quantum gate in a microfabricated surface-electrode ion trap that is operated in a cryogenic environment, and present some control techniques developed for this experiment.

We implement a Cirac-Zoller controlled-NOT (CNOT) gate using qubits represented by the atomic and motional states of a single ion. The $S\leftrightarrow D$ optical transition in $^{88}$Sr$^+$ is used as one of the qubits. The motional ground state and first excited state of the ion in the harmonic trap potential form the second qubit. The optical transition has the advantage of a long lifetime while requiring only a single laser (unlike hyperfine qubits), but the qubit is first-order Zeeman sensitive, which makes it susceptible to magnetic field noise. Taking advantage of the cryogenic environment, we stabilize the magnetic field using a pair of superconducting rings \cite{Gabrielse:91}. Since the Sr$^+$ ion qubit is not an ideal two-level system, the coupling between the sideband and carrier transitions causes level shifts known as the ac Stark shift, which must be corrected. In previous work, this has been accomplished with an additional laser field with the opposite detuning to cancel out the shift \cite{Haeffner:03}. Here, to reduce the experimental complexity of the additional acousto-optical modulators (AOMs) and optics required, Stark shift corrections are implemented in the experiment control scheme by shifting reference frames as is done in NMR \cite{Vandersypen:04}. For readout, the qubit encoded in the motional state of the ion normally cannot be measured directly, but conditional pulse sequences allow full state tomography of the qubit system. 

The control techniques developed here may be applicable to use of a single ion to probe and manipulate other systems, even though they focus on a single ion and do not necessarily imply scalability. Some such systems include the coupling of ions to superconducting qubits \cite{Tian:04}, micromechanical cantilevers \cite{Hensinger:05}, cavities \cite{Kim:09}, and wires \cite{Daniilidis:09}. In many of these experiments, maximizing the coupling requires proximity of the ion to a surface, and coherence of the motional state is also desired. 

%\subsection{paper summary}
This paper is organized as follows. The experimental setup, including the magnetic field stabilization scheme, is described in Sec. \ref{sec:expt}. Section \ref{sec:heating} briefly discusses motional state decoherence and shows that such decoherence has an insignificant effect on the gate performance in our system. Section \ref{sec:Stark} presents a theoretical model of the Stark shift correction and experimental implementation of the method. Section \ref{sec:QPT} describes the state preparation and measurement sequences that allowed us to implement quantum process tomography on the single-ion system. Section \ref{sec:CNOT} describes the realization of the CNOT gate, along with a discussion of gate performance and error sources.

\section{Experimental setup}  %%%%%%%%%%%%%%%%%%%%%%%%%
\label{sec:expt}

\subsection{Cryogenic microfabricated trap} %%%%%%%%%%%%%%%%%%%%%%%%%
The microfabricated trap is a five-rod surface-electrode design identical in geometry to that described in Ref. \cite{Labaziewicz:08}. The trap is made of niobium, and the fabrication process is similar to prior methods \cite{Labaziewicz:08} employed for gold traps, and is described briefly here. A 440-nm Nb layer is grown on a sapphire substrate by sputtering. The sheet resistance is 0.3~$\Omega$/sq at 295~K, and the superconducting transition is at $T_{c}=9.15$~K. Trap electrodes are patterned using NR9-3000 photoresist and etched using reactive ion etching with CF$_4$+O$_2$. The trap center is 100~$\mu$m above the surface. For the experiment described here, the axial and two radial trap frequencies are 2$\pi\times\{1.32, 2.4, 2.7\}$~MHz, respectively. Although a superconducting trap was used for the work described here, the effects of the superconducting material on trapping behavior will be described elsewhere \cite{Ge:09}. 

The trap is cooled and operated in a 4~K bath cryostat described in Ref.\cite{Antohi:09}. Typical ion lifetime is on the order of several hours, limited only by the liquid helium hold time. Loading is done via photoionization of a thermal vapor.

\subsection{Sr$^+$ qubit and laser system}  %%%%%%%%%%%%%%%%%%%%%%%%%

\begin{figure}
\includegraphics[width=3.3in]{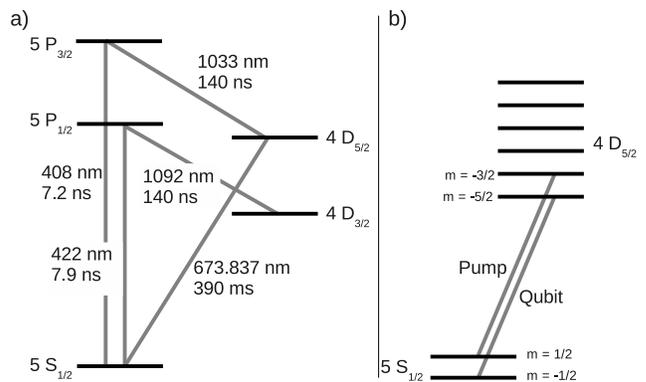}
\caption{\label{Fig:level} (a) $^{88}$Sr$^+$ level diagram. The 422- and 1091-nm transitions are used for Doppler cooling and detection. The 673.837-nm transition couples the qubit levels. (b) Details of the qubit states with Zeeman levels explicitly drawn. The ``pump'' transition is used to pump the ion out of the $S_{1/2}$($m=1/2$) state during initialization. }
\end{figure}

An atomic ion confined in a harmonic trapping potential can encode two qubits, one in its optical atomic transition and one in its lowest motional states. The $^{88}$Sr$^+$ ion has a narrow optical transition, $S_{1/2}\leftrightarrow D_{5/2}$ with a linewidth of 0.4~Hz. The $m=-1/2 \leftrightarrow m=-5/2$ levels are used for the atomic qubit transition. This transition is chosen for convenience, as along with the $P_{3/2}$ ($m=-3/2$) level it forms a closed three-level system for sideband cooling \cite{RoosThesis}. The degeneracy of the multiple Zeeman levels is lifted by applying a constant field of 4~G with external coils. To address this transition at 674~nm, a diode laser is grating-stabilized and locked to an external cavity via optical feedback \cite{Labaziewicz:07}. It is further stabilized by locking to a high-finesse cavity made of ultralow expansion glass as in Ref.\cite{Ludlow:07}. The frequency noise, indicated by the Pound-Drever-Hall error signal as measured with a spectrum analyzer, is 0.3~Hz for noise components above 1~kHz. Below 1~kHz, acoustic noise broadens the laser linewidth to $\sim$300~Hz, an estimate based on Ramsey spectroscopy measurements on the carrier $S-D$ transition assuming that the Ramsey contrast decay is caused primarily by the laser linewidth. This laser beam propagates along the axial direction of the trap, so we ignore the radial modes of motion in sideband cooling and quantum operations. Doppler cooling is performed on the $S_{1/2}$ $\leftrightarrow$ $P_{1/2}$ transition with a 422~nm diode laser. Two IR diode lasers, at 1092 and 1033~nm, repump the ion from the $D_{5/2}$ and $D_{3/2}$ states. For all measurements, the ion is initialized to the $S_{1/2}(m=-1/2)$ state and the motional ground state via a sequence of Doppler cooling, sideband cooling, and optical pumping. Figure \ref{Fig:level} shows the relevant levels of Sr$^+$ for the experiment.

A pulse sequencer \cite{PhamThesis} consisting of a field-programmable gate array (OpalKelly XEM3010-1000) and direct digital synthesis boards controls the phases, amplitudes, and lengths of the laser pulses. Switching of the beam and setting of the desired frequency and phase shift are accomplished using AOMs on the 674-, 422-, and 1033-nm lasers. Phase-coherent switching is implemented by computing the expected phase at time $t$, referenced to a fixed point in the past, for a given frequency $f$ using $\phi_0(t)=ft$(mod $2\pi$). Then, after a frequency switch at time $T$, the absolute phase of the waveform is adjusted to equal $\phi_0(T) + \phi$, where $\phi$ is any desired phase. This process allows for frequency switching while maintaining phase information throughout any arbitrary pulse sequence.

\subsection{Magnetic field stabilization}   %%%%%%%%%%%%%%%%%%%%%%%%%

When the optical qubit is encoded in a pair of levels that are first-order sensitive to magnetic fields, field fluctuations on the time scale of gate operations will decrease gate fidelity. One way of passively stabilizing the field is by use of a $\mu$-metal shield, which is expensive and inconvenient for optical access, and also mainly effective for low-frequency noise. Active stabilization of the magnetic field using a flux gate sensor and coils has been implemented in another experiment \cite{SchmidtKaler:03b}, at the cost of higher complexity. 

Superconducting solenoids have been employed for passively stabilizing ambient magnetic field fluctuations in NMR experiments, with field suppression by a factor of 156 \cite{Gabrielse:91}. A similar method for ion traps which would permit good optical access is desired. In the NMR implementation, the field needs to be stabilized over a region 1~cm in length, whereas in an ion trap the region of interest is much smaller. Our method uses the same principle of superconductive shielding, but the small region and requirement for optical access suggest a more compact approach.

\begin{figure}
\begin{tabular}{c}
\includegraphics[width=3.3in]{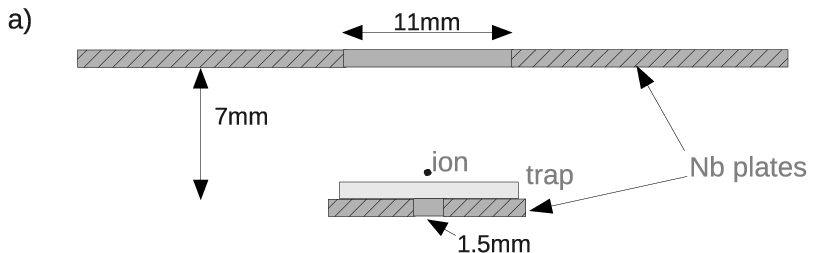} \\
\includegraphics[width=3.3in]{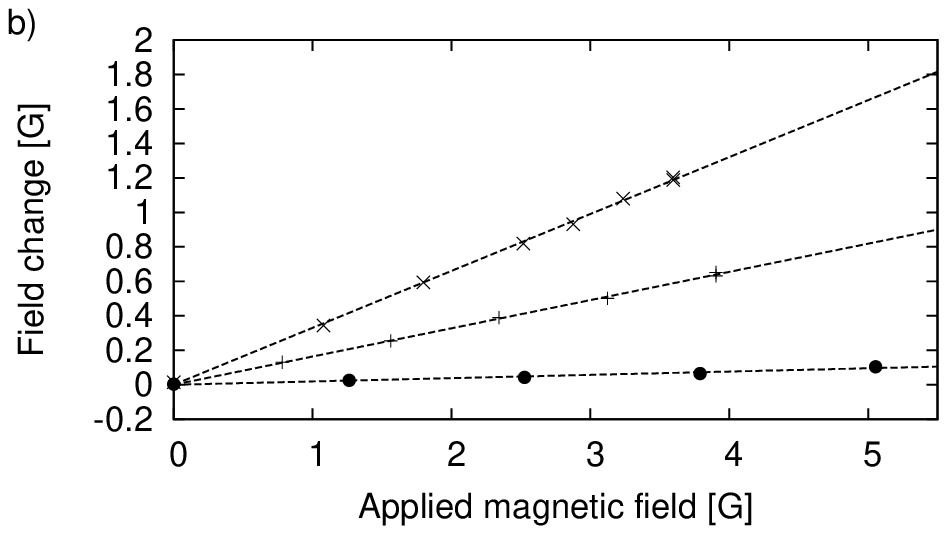} \\
\end{tabular}
\caption{\label{Fig:fieldstab} (a) Two superconducting disks, one below and one above the trapped ion, stabilize the magnetic field in the $\hat{z}$ direction. (Not to scale.) (b) Magnetic field fluctuation suppression due to the top disk only ($\times$), the bottom disk only ($+$), and both disks (\textbullet). When both disks are used, field changes are suppressed $50$-fold.}
\end{figure}

We stabilize the magnetic field by employing the persistent current in two superconducting rings, placed closely adjacent to the ion trap chip. This is a very compact and experimentally convenient arrangement, with high passive field stability and little barrier to optical access. Below the trap is a 1 $\times$ 1~cm$^2$ square Nb plate with a 1.5~mm diameter hole, located 0.5~mm below the trap center. Above the trap is a 50~cm$^2$ square plate with an 11-mm-diameter hole, located 7~mm above the trap center [Fig.\ref{Fig:fieldstab}(a)]. Both rings are 0.5~mm thick. This geometry was chosen to optimize the field suppression at the trap location using the method to calculate magnetic fields in superconducting rings described in Ref.\cite{Babaei:03}. 

With a single trapped ion we measured the field suppression by applying a constant field with external coils, cooling the trap and Nb rings to below $T_c$, and reducing the field while measuring the $S\leftrightarrow D$ transition frequency. The magnetic field is calculated from the Zeeman splitting between the $m=-1/2 \leftrightarrow m=-5/2$ transition and the $m=+1/2 \leftrightarrow m=-3/2$ transitions. A 50-fold reduction in field sensitivity was observed (Fig.\ref{Fig:fieldstab}), in agreement with the numerical calculation. To determine the effectiveness of the noise suppression on coherence of the atomic qubit, we measured the decay of Ramsey fringes as a function of the separation of the Ramsey $\pi/2$ rotations on the carrier $S\leftrightarrow D$ transition. Such a measurement also includes effects caused by laser linewidth and the drift in laser-ion distance. We found that reducing the magnetic field noise by a factor of 50 did not improve the coherence time by more than a factor of 2, from $T_2^*\sim350~\mu$s to $\sim660~\mu$s. This suggests that magnetic field noise is no longer a dominant source of decoherence when compared to laser linewidth. Although this measurement was done under dc and the dominant source of magnetic field fluctuations is frequencies near 60~Hz and its harmonics, we can estimate the bandwidth of this compensation scheme by relating it to material properties of niobium as a type-II superconductor. The field suppression factor is determined by how fast the induced currents in the superconducting rings respond to changes in the external field, which depends on the ring's inductance (a geometric factor independent of frequency) and resistance. Above the first critical field, type-II superconductors exhibit flux pinning, which leads to ac resistance, but the critical field for niobium is on the order of 1000~G \cite{Poole:book, Beall:69}. Below the critical field, superconductors can still exhibit a frequency-dependent AC resistance as described in Ref. \cite{Kaxiras:book}. However, for niobium the effect is not significant until frequencies up to $\sim$$10^{12}$~Hz. Therefore at typical bias fields (4~G) and frequencies relevant to our qubit ($<$1~kHz), niobium behaves as a perfect superconductor and we expect the field suppression factor to be the same as that measured under dc. 

Greater reduction can be obtained by optimizing the geometry further, for example, by decreasing the distance between the plates to 4~mm, but is not implemented because of physical constraints in the apparatus. This method stabilizes the magnetic field only along the axis of the superconducting rings, but since the 4-G bias field defining the quantization axis is applied in the same direction, field noise in the x or y direction contributes only quadratically to the change in the total field \cite{Gabrielse:91}. 

\section{Motional state coherence}   %%%%%%%%%%%%%%%%%%%%%%%%%
\label{sec:heating}

The Cirac-Zoller CNOT gate employs superpositions of ion motional states as intermediate states during the gate, and thus is sensitive to motional decoherence. In particular, a high ion heating rate will reduce the gate fidelity. An upper bound on the maximum heating rate tolerable, $\dot{n}_{\rm max}$, can be given by consideration of the total time $T_{\rm gate}$ required for the pulse sequence implementing the CNOT gate, together with a design goal for the gate error probability $p_{\rm gate}$ desired.  Assuming that a single quantum of change due to heating will cause a gate error, then $\dot{n}_{\rm max} < T_{\rm gate} / p_{\rm gate}$. For $T_{\rm gate}\sim 230$ $\mu$s (for our experiment), a heating rate of $\dot{n}_{\rm max} < 40$ quanta/s is needed to get $p_{\rm gate}\sim 0.01$.

\begin{figure}
\includegraphics[width=3.3in]{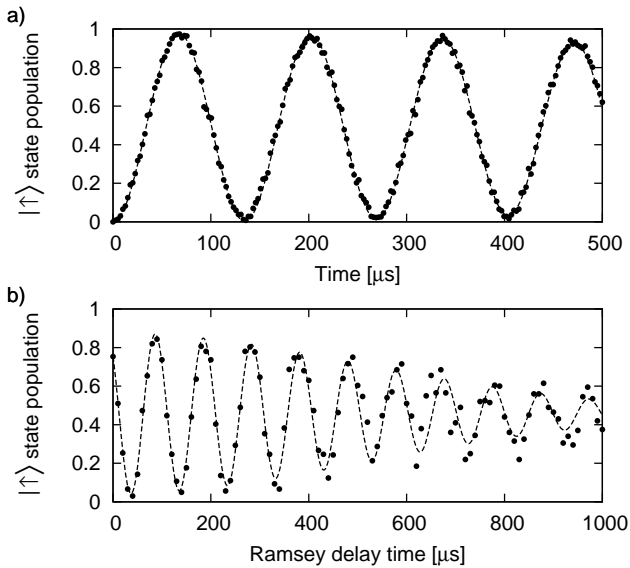}
\caption{\label{Fig:BSB} (a) Rabi oscillations on the blue sideband. The fitted initial contrast is 97.6(3)\% and the frequency is 46.7~kHz. (b) Ramsey spectroscopy on the blue sideband. The fitted Gaussian envelope of the decay has time constant $T_2^* = 622(37) \mu$s.}
\end{figure}

We measured the heating rate of the trap at the operating secular frequency of $2\pi~\times~$1.32~MHz. The number of motional quanta is measured by probing the blue and red sidebands of the $S\leftrightarrow D$ transition using the shelving technique, and comparing the ratio of shelving probability on each sideband \cite{Turchette:00}. The heating rate is determined by varying the delay before readout and comparing the number of quanta versus delay time. The measured heating rate is weakly dependent on the rf voltage and dc compensation voltages. Noise on the rf pseudopotential can cause heating \cite{Wineland:98, Blakestad:09}, so the ion micromotion is minimized using the photon correlation method \cite{Berkeland:98}. For more details about the measurements, see Ref. \cite{Labaziewicz:08d}. The heating rate can also depend on the trap's processing history and may vary between temperature cycles \cite{Labaziewicz:08b}; for this trap, the variation is small. In a typical experimental run, the rf voltage and dc compensation values are adjusted to minimize the heating rate before the coherence time and quantum gate data are taken. Typical heating rates obtained in this trap are 4-6~quanta/s, while the lowest heating rate measured is 2.1(3)~quanta/s. Figure \ref{Fig:BSB}(a) shows Rabi flops on the blue sideband after the ion is initialized to the motional ground state with average number of quanta $\bar{n} < 0.01$. The fitted initial contrast is 97.6(3)\% and the frequency is 46.7~kHz. Motional state coherence is demonstrated by performing Ramsey spectroscopy on the blue sideband [Fig.\ref{Fig:BSB}(b)]. The coherence time $T_2^*$ is 622(37)~$\mu$s. This is comparable to the coherence time of 660(12)~$\mu$s of the atomic qubit as measured by the same method on the carrier transition.% data is 'HT_RF190_31.8DC_end25' from Dec15

\section{Stark shift correction}  %%%%%%%%%%%%%%%%%%%%%%%%
\label{sec:Stark}

When a two-level atom encoding a qubit is driven off resonance, as on a sideband transition, it excites the carrier transition and creates an ac Stark shift. In a real ion with multiple levels, additional complication comes from other transitions that contribute shifts which are independent of the laser detuning. In the past, correction for the Stark shift has been done by using an additional laser detuned to the opposite sideband transition to cancel the shift \cite{Haeffner:03}. In qubits addressed by a Raman transition, this can also be accomplished by changing the power ratio of the Raman pulses \cite{Haze:09}. In this work, the Stark shift correction is done by calculating the shift and accounting for it in the pulse sequencer, following an example in NMR \cite{Vandersypen:04}. Here, we develop a systematic model of the light shifts experienced by a single trapped ion. 

The Stark shift is traditionally a phase shift caused by a small change in the transition frequency caused by level shifts. In reality, it is a unitary transform involving more than just a change of energy levels. We also take this into consideration later as a ``generalized'' Stark shift. The model presented here is adapted from well-known methods in NMR and included for pedagogical reasons. Section \ref{sec:Starkion} identifies the reference frames useful for discussing the single ion in the context of quantum control. The generalized Stark shift correction operation is then derived as a result of switching between these frames. In Sec. \ref{sec:Starkcorr} we apply this method to our single-ion system and describe how to calculate the appropriate Stark shift correction for any gate in an arbitrary gate sequence. Section \ref{sec:Starkresult} describes the measurement of the ac Stark shift and results of the Stark shift correction.

\subsection{Stark shift on carrier: Simple free-ion model}   %%%%%%%%%%%%%%%%%%%%%%%%
\label{sec:Starkion}

There are several useful frames of reference to describe the two-level atom model. Consider a single ion at a fixed position in free space, interacting with a single-mode laser. Let this be described by the laboratory reference frame Hamiltonian (with the rotating-wave approximation)
\be    H_0 = \omega_0 \sigma_z + \Omega\sigma_x \cos \omega t + \Omega\sigma_y \sin \omega t
\ee
where $\omega_0$ is the optical transition frequency, $\sigma_x$, $\sigma_y$, $\sigma_z$ are spin-1/2 operators corresponding to the Pauli matrices with eigenvalues $\pm 1/2$, and $\Omega$ is the Rabi frequency.

Let the laser be applied at frequency $\omega = \omega_0 + \delta$, such that we may define
\be    H_L = \omega \sigma_z 
\ee
as a convenient frame of reference. In the frame of the laser, the Hamiltonian is 
\be    V_L = -\delta \sigma_z + \Omega \sigma_x .
\ee

The frame of reference we wish to use for quantum computation (QC frame) is defined by the Hamiltonian
\be    H_{\rm QC} = \omega_0 \sigma_z .
\ee
Thus, if we define a state in this frame as
\be    |\gamma(t)\rangle = e^{+iH_{\rm QC} t}|\psi(t)\rangle 
\ee
where $|\psi(t)\rangle$ is the state in the laboratory frame, then we find that
\be    |\gamma(t)\rangle = e^{i\delta \sigma_z t} e^{-i (\delta \sigma_z + \Omega\sigma_x)t} |\gamma(0)\rangle 
\ee
assuming that $|\gamma(0)\rangle = |\psi(0)\rangle$. 

The generalized Stark shift correction operation that needs to be applied is thus $R^\dagger$, where
\be
    R = e^{i\delta \sigma_z t} e^{-i (\delta \sigma_z + \Omega\sigma_x)t} .
\ee
This is an operator that rotates about an axis 
\be
    \hat n = \frac{\hat z + (\Omega/\delta)\hat x}{\sqrt{1+ (\Omega/\delta) ^2}}.
\ee
When the detuning is very large compared with the Rabi frequency, the maximum rotation about the $\hat{x}$ axis, which corresponds to a population change, can be bounded by $\Omega^2 / \delta^2$ for a $\pi$~rotation about $\hat{n}$. For our experimental parameters (Sec. \ref{sec:gateperf}), this is less than 1\%. Therefore the Stark shift is traditionally approximated as a rotation about the $\hat{z}$ axis, $R_z(\theta) = e^{i\theta\sigma_z}$. We can compute what this operation and the rotation angle $\theta$ would be by looking for the $R_z$ closest to $R$. The angle of rotation of the operator $e^{-i (\delta \sigma_z + \Omega\sigma_x)t}$ is $\sqrt{\delta^2 + \Omega^2}t$, while the angle of rotation of the operator $e^{i\delta \sigma_z t}$ is $\delta t$. Thus, if one ignored the axes of rotation and treated the first operator as if it were also a rotation about $\hat{z}$, then the Stark shift correction would be a rotation by angle
\be
    \left[ \delta - \sqrt{\delta^2 + \Omega^2} \right] t
\ee
about $\hat{z}$. 

\subsection{Stark shift corrections for arbitrary gate sequences}   %%%%%%%%%%%%%%%%%%%%%
\label{sec:Starkcorr}

We now examine a real experimental situation with a multilevel ion. To verify our proposed Stark shift correction and later to consider the effect of error sources on gate fidelity, we simulated gate operations by modeling the action of lasers on the full system Hamiltonian in the space formed by $\{|D\rangle,|S\rangle\}\otimes\{|0\rangle,|1\rangle,|2\rangle\}$. 

Exact simulation of the action of the lasers on the computational space requires the use of both laser and QC frames. The full Hamiltonian is time independent in the laser frame, suggesting that gate operations should be computed in that frame. The states used in quantum computation are defined in the QC frame, where they are stationary without the interaction applied. Simulation of a gate sequence will therefore require frequent switching between the frames, for which we define the operator $U_{\rm LQC}(t)$. 

Computation is performed by moving to the laser frame, exponentiating $V_{L}$, and moving back to the QC frame.
For example, a gate performed by application of a laser pulse of detuning $\delta$, phase $\phi$, starting at time $t_0$ for time $t$, can be computed in the QC frame to be
\be
U(\delta, \phi, t, t_0) = U_{\rm LQC}(t+t_0) e^{-\imath V_L t/\hbar} U_{\rm LQC}(t_0)^\dagger
\ee
where $V_L$ and $U_{\rm LQC}$ depend on the laser detuning, phase, and Rabi frequency as well as the trap parameters.
Note that each laser detuning and trap frequency define a separate laser frame.
The operator $U_{\rm LQC}$ moves between the unique QC frame and one of the infinite number of laser frames.

Let $U_\phi(\phi) = e^{-i\phi}$ be a phase shift on the $D$ states:
\be
U_\phi(\phi) = \left( \begin{array}{cc}
e^{i \phi} & 0 \\
0 & 1 \\ 
\end{array}\right) \otimes I_3,
\ee
where $I_n$ is the identity matrix of size $n\times n$. Here the 2$\times$2 matrix acts on $\{|D\rangle,|S\rangle\}$ and $I_3$ acts on the motional states. Experimentally, this operator is equivalent to shifting the laser phase by $\phi$. In a sequence of gates, application of such a phase rotation implies shifting the laser phases of all subsequent gates. 

In a multilevel atom, there are other transitions that are off-resonantly coupled to the laser and contribute to additional phase shifts that are detuning independent. In our modeling of the Sr$^+$ computation presented here, we include the $S_{1/2} \leftrightarrow P_{1/2}$, $S_{1/2} \leftrightarrow P_{3/2}$, and $D_{1/2} \leftrightarrow P_{3/2}$ transitions. The matrix elements of all these transitions, which determine the resulting shift, can be calculated as in Ref.\cite{James:98a}.

For gates performed on the carrier transition, since the duration of carrier gates is shorter than that of sideband gates by the Lamb-Dicke factor $\eta$ ($=0.06$ for our case), carrier gates take only a small fraction of the total time in a typical gate sequence ($\sim$2\% in the CNOT pulse sequence). Thus we ignore off-resonant coupling to the motional sidebands and coupling to far-off-resonant transitions. For gates on the sideband transitions, consider an interaction with laser detuning $\d \gg \Om$, carrier Rabi frequency $\Omega$, and phase $\phi$ applied for time $t$ starting at time $t_0$. There are three separate phase shifts that need to be cancelled:
\begin{itemize}
\item Stark shift of the ground and excited states can be removed by rotating the phase of the $\ket{e}$ state by $\phi_s = -\left(\d - \sqrt{\d^2 + \Om^2}\right)t$, equivalent to applying $e^{-\imath \sigma_z \phi_s}$, following the offending gate. $Z$ rotations can be performed by changing the phases of all subsequent laser pulses by $\phi_s$.

\item Resonant excitation of sidebands is applied at a frequency Stark-shifted owing to the carrier. The laser frame corresponding to that frequency will rotate with respect to the unshifted states at a rate proportional to the Stark shift. To bring the laser frame and unshifted states in phase, the laser phase has to be shifted by $\phi_f = \left(\d - \sqrt{\d^2 + \Om^2} \right)t_0$. Such a phase shift is equivalent to applying $e^{-\imath \sigma_z \phi_f}$ before the gate, and $e^{\imath \sigma_z \phi_f}$ after.

\item Off-resonant phase shifts account for approximately 10\% of the total Stark shift in Sr$^+$. Let $\Delta_0$ be a constant factor to account for these off-resonant phase shifts.
\end{itemize}

Define the carrier gate $U_c$, sideband gate $U_m$, and phase correction $\Delta$ as follows. Along with the gate time $t$ and gate starting time $t_0$, these variables contain all the information relevant to calculating the required Stark shift correction.
\begin{eqnarray}
    U_c &=& U \\
    U_m &=& U_{\phi}\left(-\Delta(t+t_0)\right) U U_{\phi}\left(\Delta t_0\right) \\
    \Delta &=& \left[ \delta - \sqrt{\delta^2 + \Omega^2} \right] + \Delta_0 
\end{eqnarray}

From the definition of $U$ and properties of the exponential function, it can be shown that the phase correction on the $n$th gate $U^n$ in an arbitrary gate sequence is
\be
U^n = U_{\phi}\left(-\sum_{i=1}^{n} \Delta^i t^i\right) U\left(\delta, \phi+\Delta^n t_0^n - \sum_{i=1}^{n-1} \Delta^i t^i, t^n, t_0^n\right) . 
\ee
This phase correction consists of the appropriate correction for that particular gate plus a global phase, the sum of all phase corrections applied to previous gates. In our pulse sequencer \cite{PhamThesis}, the global time and global phase are kept as internal registers, and are used to calculate the appropriate phase correction every time the qubit laser phase is set during a pulse sequence. 

\subsection{Results}  %%%%%%%%%%%%%%%%%%%%%%
\label{sec:Starkresult}

\begin{figure}
\begin{tabular}{c}
\includegraphics[width=3.3in]{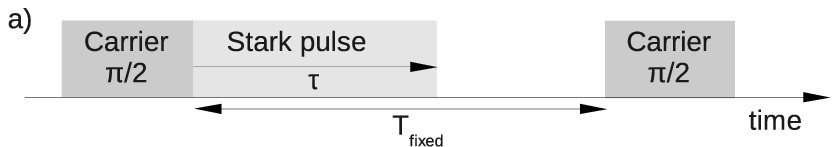}\\
\includegraphics[width=3.3in]{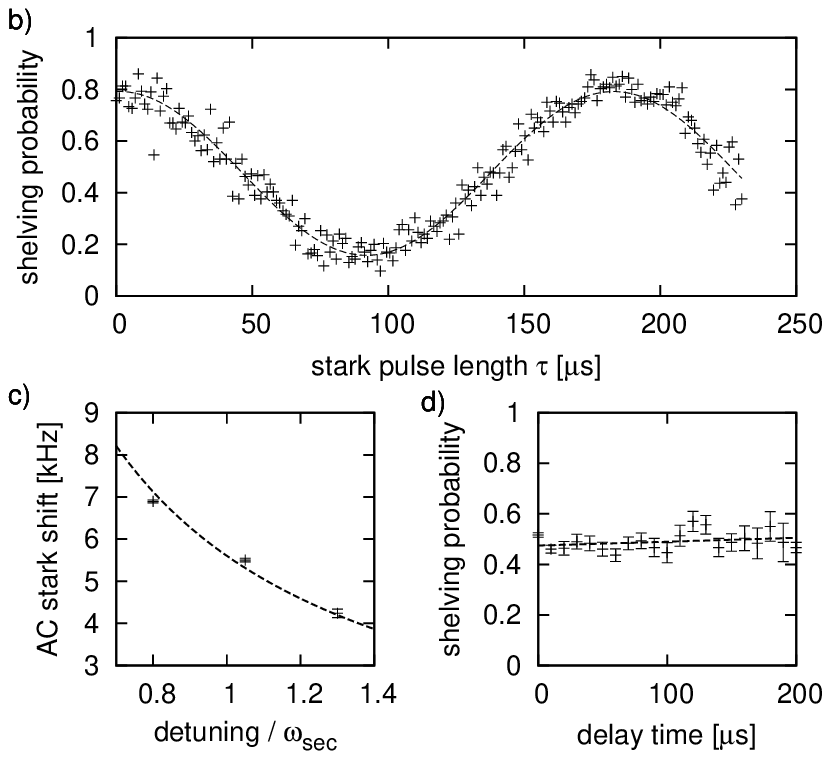}\\
\end{tabular}
\caption{\label{Fig:Stark} (a) Pulse sequence used to measure ac Stark shift, from Ref.\cite{Haeffner:03}. The delay between two carrier $\pi/2$ pulses is $T_{\rm fixed}=230 \mu$s while the length $\tau$ of the Stark pulse is varied. (b) Typical measurement with the pulse sequence in (a). The ac Stark shift is given by the oscillation frequency of the shelving probability. Here, the fixed laser detuning is 1.05$\omega_{\rm sec}$ which gives an ac Stark shift of $2\pi \times 5.50(4)$~kHz. (c) Measured ac Stark shift as a function of detuning in units of $\omega_{\rm sec}$, fitted to the Stark shift model with the detuning-independent shift as a free parameter. (d) Ramsey spectroscopy on the sideband, demonstrating compensation of Stark shift.}
\end{figure}

Ramsey spectroscopy on the blue sideband can be used to characterize the effectiveness of the Stark shift correction. Using the methods described in Ref.\cite{Haeffner:03}, we measure the ac Stark shift for various detunings and compare to the theoretical model. Figure \ref{Fig:Stark}(b) shows the typical oscillation in the shelving probability $P(D)$ when a pulse detuned from the S$\leftrightarrow$D transition (Stark pulse) of varying duration is applied. The ac Stark shift is given by the oscillation frequency. This shift is measured for several values of detuning and is shown in Fig.\ref{Fig:Stark}(c) along with a one-parameter fit to $A/x+b$, where the fixed parameter is $A = \Omega^2/(2\omega_{\rm sec})$, the detuning-dependent shift to first order, and $b$ is the detuning-independent Stark shift caused by farther-off-resonant transitions. The fitted offset is $b = -2\pi \times 0.5(1)$~kHz, in agreement with $-2\pi \times0.50$~kHz predicted by theory. 

The effectiveness of AC Stark shift compensation was evaluated by performing Ramsey spectroscopy on the sideband and varying the delay time between the two pulses, with both Ramsey pulses shifted by $\pi/2$. In the absence of uncorrected Stark shifts, the expected $P(D)$ is 1/2 for all delay times. Figure \ref{Fig:Stark}(d) shows the result of such a measurement. Here, the secular frequency is determined by taking a spectrum and fitting to the sideband; then the Ramsey sequence is performed. The parameter $\phi_s$ is fixed in the experiment control hardware while $\Delta_0$ is tuned such that $P(D)$ is maintained near 1/2. From the slope of Fig.\ref{Fig:Stark}(d), we estimate the residual Stark shift to be $2\pi \times 24(20)$ Hz.
% fitted to slope and use slope=w where w=2pi*ACshift

\section{Quantum process tomography on a single ion}    %%%%%%%%%%%%%%%%%%%%%%%%
\label{sec:QPT}

With $N=2$ qubits encoded in a single ion and methods of coupling and controlling these states, a Cirac-Zoller CNOT gate can be implemented \cite{Cirac:95}. The CNOT gate is universal in that all quantum operations can be decomposed into single-qubit operations and the CNOT gate, and is thus of interest for implementing quantum information processing in ion traps. To evaluate the performance of such a gate, we prepare the system in a set of basis states that spans the space of $2^N\times2^N$ density matrices and perform a set of measurements that completely specifies the resulting state (state tomography). Quantum process tomography (QPT) is performed on the two qubits to construct the process matrix, allowing a full characterization of the gate. Section \ref{sec:QPTion} gives a brief summary of state tomography using conditional measurements. Section \ref{sec:QPTdef} describes a minimal set of available measurements and operations in this two-qubit single-ion system necessary for QPT. Sections \ref{sec:QPTprep} and \ref{sec:QPTmeas} list the pulse sequences for preparing all basis states and measuring the outcome. Section \ref{sec:QPTresult} briefly describes the construction of the process matrix that fully characterizes the gate from these measurements.

\subsection{Two-qubit state tomography for one ion}   %%%%%%%%%%%%%%%%%%%%%%%%%%
\label{sec:QPTion}

State tomography on the single-ion system of atomic and motional qubits requires a nontrivial set of operations, since a single qubit rotation on the motional qubit cannot be realized directly except by first swapping it with the internal state, performing the desired gate, then swapping back. The swap operation is complicated since the most straightforward physical operations, red- and blue-sideband pulses, generally take the system out of the computational space, and into higher-order motional states such as $|2\rangle$ \cite{Childs:00}. For the CNOT gate, a set of composite pulse sequences can keep the system in the computational space. But if the goal is measurement of the two-qubit state space rather than the realization of a coherent operation, an alternative approach can be employed. A sequence of measurements, with the second conditioned on the results of the first, can suffice to allow full state tomography on the two-qubit atomic+motional state space. This is an extension of the single-ion tomography technique described in \cite{Monroe:95}.

The conditional measurement sequence is as follows. First we apply an optional $\pi$ pulse on the carrier transition; then the internal atomic state is measured by fluorescence detection. When this measurement scatters photons, it provides information about the internal state only and the motional state information is lost. When this first measurement does not scatter photons, a $\pi$ pulse is applied on the blue-sideband transition, which allows measurement of the population in the state pairs $\{|S0\rangle,|S1\rangle\}$ or $\{|D0\rangle,|D1\rangle\}$, depending on whether the initial carrier $\pi$ pulse was applied or not. Two measurements, with and without the carrier pulse, are sufficient to determine the population in all four states.

\subsection{Process tomography: Operator definitions}
\label{sec:QPTdef}

The state tomographic measurement just described measures state populations only, which are the diagonal elements of the full density matrix. Relative phases between qubit states, which determine coherence properties of the state, are also needed in order to perform complete process tomography. The phases can be obtained by appropriate rotations of the qubits prior to measurement. Here we define the measurement and rotation operators for the sections following.

The single available measurement is the usual fluorescence detection, which is a projective measurement into the $|S\rangle$ state, denoted $P_S$. Let $P_D$ denote a projection into the $\ket{D}$ state. The matrices for $P_S$ and $P_D$ in the basis $\ket{D0, D1, D2, S0, S1, S2}$ are
\be
P_S = \left( \begin{array}{cc}
0 & 0 \\
0 & 1 \\ 
\end{array}\right) \otimes I_3. \quad
P_D = I_6 - P_S
\ee.

The available unitary operations are as follows
\begin{itemize}
\item $R_x(\theta)$, $R_y(\theta)$: Single qubit (carrier) rotations on the $\{|S\rangle,|D\rangle\}$ qubit.
\item $R_x^+(\theta)$, $R_y^+(\theta)$: Blue-sideband rotations, connecting $\{|S0\rangle,|D1\rangle\}$ and $\{|S1\rangle,|D2\rangle\}$ (neglecting higher-order vibrational modes). $\theta$ is the rotation angle on the $\{|S0\rangle,|D1\rangle\}$ manifold.
\item Red-sideband rotations can be defined similarly, but are actually not necessary for construction of a complete measurement set.
\end{itemize}
Explicitly, these rotation matrices are defined as follows:
\begin{eqnarray}
R_x(\theta) &=& \exp[-i \theta (\sigma_x \otimes I_3 )] \nonumber \\
R_y(\theta) &=& \exp[-i \theta (\sigma_y \otimes I_3 )] \nonumber \\
R_x^+(\theta) &=& \exp[\theta ( \sigma_+ \otimes a^\dagger - \sigma_- \otimes a) /2 ] \nonumber \\
R_y^+(\theta) &=& \exp[-i \theta (\sigma_+ \otimes a^\dagger + \sigma_- \otimes a) /2 ]
\end{eqnarray}
where $a^+$ and $a$ are the creation and annilation operators in the Jaynes-Cummings Hamiltonian.

\subsection{State preparation} %%%%%%%%%%%%%%%%%%%
\label{sec:QPTprep}

For every measurement sequence, the ion is initialized to the state $\Psi_0 \equiv \ket{S0}$. The sequences of operations listed in Table \ref{Tbl:StatePrep} generates the 16 input states that span the space of 4$\times$4 density matrices created from the product states $\ket{D0, D1, S0, S1}$.

\begin{table}
\begin{tabular}{llc}
\hline
 $\Psi(i)$     & Operations applied to $\Psi_0$  &  State \\
\hline
$\Psi(1)$  & $R_y(-\pi)                      $ & $\ket{D0}$ \\
$\Psi(2)$  & $R_x^+(-\pi)                     $ & $\ket{D1}$  \\
$\Psi(3)$  & $I                               $ & $\ket{S0}$  \\
$\Psi(4)$  & $R_y(\pi) R_x^+(\pi)         $ & $\ket{S1}$  \\
$\Psi(5)$  & $R_x^+(\pi) R_y(-\pi/2)        $ & $(\ket{D0}+\ket{D1})/\sqrt2 $ \\
$\Psi(6)$  & $R_y^+(-\pi) R_y(-\pi/2)      $ & $(\ket{D0}+i\ket{D1})/\sqrt2 $ \\
$\Psi(7)$  & $R_y(-\pi/2)                     $ & $(\ket{D0}+\ket{S0})/\sqrt2 $ \\
$\Psi(8)$  & $R_x(\pi/2)                     $ & $(\ket{D0}+i\ket{S0})/\sqrt2 $ \\
$\Psi(9)$  & $R_y(-\pi) R_x^+(-\pi/2)        $ & $(\ket{D0}+\ket{S1})/\sqrt2  $\\
$\Psi(10)$ & $R_y(-\pi) R_x^+(\pi/2)    $ & $(\ket{D0}+i\ket{S1})/\sqrt2 $ \\
$\Psi(11)$ & $R_x^+(\pi/2)                   $ & $(\ket{D1}+\ket{S0})/\sqrt2  $ \\
$\Psi(12)$ & $R_y^+(\pi/2)                    $ & $(\ket{D1}+i\ket{S0})/\sqrt2 $  \\
$\Psi(13)$ & $R_y(\pi/2) R_x^+(\pi)          $ & $(\ket{D1}+\ket{S1})/\sqrt2  $\\
$\Psi(14)$ & $R_x(-\pi/2) R_x^+(-\pi)        $ & $(\ket{D1}+i\ket{S1})/\sqrt2 $ \\
$\Psi(15)$ & $R_y(-\pi) R_x^+(-\pi) R_y(\pi/2)  $ & $(\ket{S0}+\ket{S1})/\sqrt2 $ \\
$\Psi(16)$ & $R_y(-\pi) R_y^+(\pi) R_y(\pi/2)   $ & $(\ket{S0}+i\ket{S1})/\sqrt2 $ \\
\hline
\end{tabular}
\caption{\label{Tbl:StatePrep} State preparation operations.}
\end{table}

\subsection{Complete basis of measurements} %%%%%%%%%%%%%%%%%%%
\label{sec:QPTmeas}

The following is a procedure for performing complete state tomography of the two-qubit $\{|S\rangle,|D\rangle\}\otimes\{|0\rangle,|1\rangle\}$ state of a single ion, using the measurements and operations in Sec \ref{sec:QPTdef}. This is a generalization of the method used to measure just the diagonal elements of the density matrix. There are two kinds of measurement used; we call them $M_U$ and $M_{UV}$.

\textbf{$M_U$} involves performing a unitary operation $U$ on the input state and then projecting into the $|S\rangle$ subspace $P_S$. This is described by the measurement operator
\be    M_U(U) = U^\dagger P_S U 
\ee
Typically, $U$ will be a rotation in the $\{|S\rangle,|D\rangle\}$ subspace, implemented by a carrier transition pulse.

\textbf{$M_{UV}$} involves first performing a unitary operation $U$ on the input state and making a measurement to detect fluorescence, which is equivalent to projecting to the $\ket{S}$ subspace. Since $\ket{D}$ is long lived, this projection leaves the $\{|D0\rangle,|D1\rangle,\ldots\}$ subspace undisturbed, but motional state information is lost if the ion is in state $\ket{S,n}$. If no fluorescence is detected, the postmeasurement state is $P_D\rho P_D$. Conditioned on the first measurement returning $\ket{D}$ (no fluorescence), a unitary transform $V$ is performed, and finally another into the $|S\rangle$ subspace $P_S$. If the first measurement returns fluorescence, the measurement sequence stops, in which case only information about the atomic state is obtained. $M_{UV}$ is described by the measurement operator
\be    M_{UV}(U,V) = U^\dagger P_D V^\dagger P_S V P_D U 
\ee
Typically, $U$ will be a rotation in the $\{|S\rangle,|D\rangle\}$ subspace, while $V$ will be one or more rotations on the carrier and the red or blue sideband. 

The measurements listed in Table \ref{Tbl:StateMeas} provide a complete basis of observables from which the full density matrix $\rho$ can be reconstructed, assuming that $\rho$ is initially in only the two-qubit computational subspace. These measurement observables are linearly independent. 

\begin{table}[htb]
\begin{tabular}{llc}
\hline
 $M_j$  & Measurement functions \\
\hline
 $M_1$ & $M_U$($I$) \\
 $M_2$ & $M_{UV}$($I$, $R_y^+(\pi)$) \\
 $M_3$ & $M_{UV}$($R_y(\pi)$, $R_y^+(\pi)$) \\
 $M_4$ & $M_U$($R_y(\pi/2)$) \\
 $M_5$ & $M_U$($R_x(\pi/2)$) \\
 $M_6$ & $M_{UV}$($I$, $R_y(\pi/2) R_y^+(\pi/2)$) \\
 $M_7$ & $M_{UV}$($R_y(\pi)$, $R_y(\pi/2) R_y^+(\pi/2)$)  \\
 $M_8$ & $M_{UV}$($R_y(\pi/2)$, $R_y(\pi/2) R_y^+(\pi/2)$) \\
 $M_9$ & $M_{UV}$($R_y(\pi/2)$, $R_x(\pi/2) R_y^+(\pi/2)$) \\
 $M_{10}$ & $M_{UV}$($I$, $R_x(\pi/2) R_y^+(\pi/2)$) \\
 $M_{11}$ & $M_{UV}$($R_x(\pi)$, $R_x(\pi/2) R_y^+(\pi/2)$) \\
 $M_{12}$ & $M_{UV}$($R_x(\pi/2)$, $R_x(\pi/2) R_y^+(\pi/2)$) \\
 $M_{13}$ & $M_{UV}$($R_x(\pi/2)$, $R_y(\pi/2) R_y^+(\pi/2)$) \\
 $M_{14}$ & $M_{UV}$($R_y(\pi/2)$, $R_y^+(\pi/2)$) \\
 $M_{15}$ \quad & $M_{UV}$($R_x(\pi/2)$, $R_y^+(\pi/2)$) \\
\hline
\end{tabular}
\caption{\label{Tbl:StateMeas} State measurement functions.}
\end{table}

The relationship between measurements and the density matrix can be expressed by a matrix $A$ with elements
\be
A_{ij} = M_j(\Psi(i)).
\ee
The full density matrix $\rho$ can be reconstructed as:
\be
\rho = \sum_{ij} m_j A_{ij}^{-1} \ket{\Psi_i}\bra{\Psi_i}
\ee
where $m_j$ is the result of measurement $M_j$.

\subsection{Construction of the process matrix}  %%%%%%%%%%%%%%%%%%%%%%%%%
\label{sec:QPTresult}

A quantum gate including all error sources can be represented by the operation $\mathcal{E} (\rho)$, which can be written in the operator sum representation as 
\begin{equation}
\mathcal{E}(\rho) = \sum_{mn} E_m \rho E_n^{\dagger} \chi_{mn}
\end{equation}
where $\rho$ is the input state and $E_i$ is a basis of the set of operators on the state space. The process matrix $\chi_{mn}$ contains the full gate information. For two qubits, the state space is spanned by 16 basis states, and $16^2$ elements define the $\chi$-matrix, although it only has 16$\times$15 independent degrees of freedom because of normalization. This is reflected in the fact that only 15 measurements are needed. The $\chi$ matrix can be obtained by inverting the above relation. To avoid unphysical results (namely, a non-positive-semidefinite $\rho$, Tr$(\rho^2)>1$) caused by statistical quantum error in the experiment, a maximum-likelihood estimation algorithm \cite{James:01} is employed to determine the physical operation $\mathcal{E}$ that most likely generated the measured data. An alternate, iterative algorithm is presented in Ref.\cite{Jezek:03}.

\section{Single-ion CNOT gate} %%%%%%%%%%%%%%%%%%%%%%%%
\label{sec:CNOT}

The CNOT gate is implemented with the pulse sequence described in Ref.\cite{SchmidtKaler:03}. The optical transition is the control qubit, and the motional ground and first excited states are used as the target qubit. In the product basis $\{\ket{D0}, \ket{D1}, \ket{S0}, \ket{S1}\}$, the unitary matrix implemented is 
\be
\begin{array}{rcl}
U & = & \left( \begin{array}{cccc}
1 & 0 & 0 & 0 \\
0 & -1 & 0 & 0 \\
0 & 0 & 0 & 1 \\
0 & 0 & -1 & 0 \end{array}\right) \\
& = & \frac{1}{2} (-iY\otimes Z + Z\otimes I + Z\otimes Z + iY \otimes Z). 
\end{array}
\ee
This differs from the ideal CNOT matrix by only single-qubit phase shifts. Section \ref{sec:gateperf} describes the achieved gate fidelities and Sec. \ref{sec:gateerr} discusses the major known error sources that compromise gate fidelity.

\subsection{Gate performance}  %%%%%%%%%%%%%%%%%%%%%
\label{sec:gateperf}

\begin{figure*}
\includegraphics{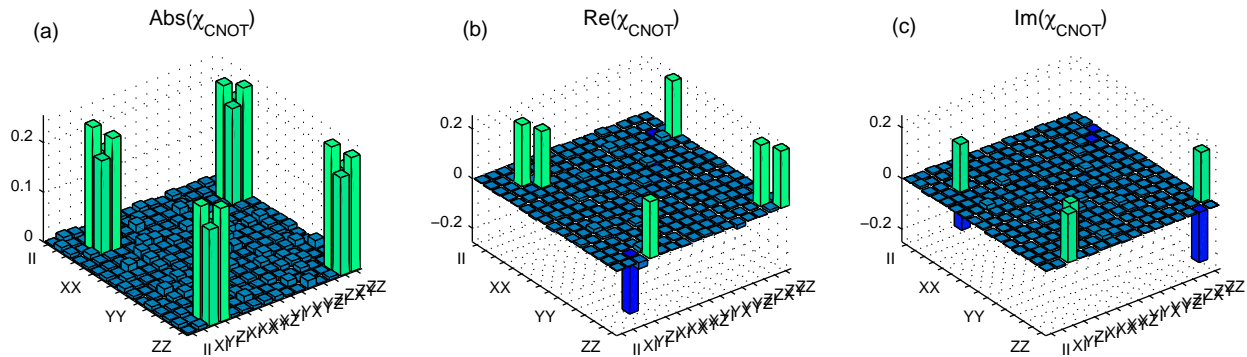}
\caption{\label{Fig:PTCNOT} Process tomography on the CNOT gate. (a), (b), and (c) show the absolute, real, and imaginary parts of the $\chi$ matrix, respectively.}
\end{figure*}

Quantum process tomography was carried out to evaluate the performance of various gates on the two qubits of a single ion. The ion in its motional and atomic ground state is initialized to one of the 16 input states in Table \ref{Tbl:StatePrep}. Then the gate is applied, and the output state is determined by making all of the measurements listed in Table \ref{Tbl:StateMeas}. The longest duration of the full measurement sequence (excluding the gate) is 610~$\mu$s, and a single CNOT gate takes 230~$\mu$s. These durations are determined by the Rabi frequency on the carrier $\Omega$ = 2$\pi \times$125~kHz and on the sideband $\Omega_{\rm BSB}$ = 2$\pi \times$7.7~kHz, and the secular frequency $\omega_{\rm sec}$ = 2$\pi \times$1.32~MHz. The resulting $\chi$ matrix for the CNOT gate is shown in Fig.\ref{Fig:PTCNOT}.

We evaluate the performance of the identity gate (all preparation and measurement sequences performed with no gate in between), the single CNOT gate, and two concatenated CNOT gates (CNOTx2). The results are shown in Table \ref{Tbl:Data}. The process fidelity is defined as $F_p =$ Tr$(\chi_{\rm id}\chi_{\rm expt})$, where $\chi_{\rm id}$ is the ideal $\chi$ matrix calculated with the ideal unitary operation $U$, and $\chi_{\rm expt}$ is experimentally obtained using maximum-likelihood estimation. We also calculate the mean fidelity $F_{\rm mean}$, based on the overlap between the expected and measured density matrices, Tr$(\rho_{\rm id}\rho_{\rm expt})$, averaged over all prepared and measured basis states, as in Ref.\cite{OBrien:04}. $F_p$ characterizes the process matrix whereas $F_{\rm mean}$ is a more direct measure of the gate performance. There exists a simple relationship between the two measures, $F_{\rm mean} = (dF_p+1)/(d+1)$ \cite{Gilchrist:05}, which is consistent with the independently calculated values for our data. Error bars on $F_p$ are calculated from quantum projection noise using Monte Carlo methods \cite{Roos:04}. The large error bars on $F_{\rm mean}$ occur because certain measured basis states consistently have a higher or lower overlap with the ideal states. In general, states that involve multiple pulses to create entanglement are more susceptible to error and therefore have a lower fidelity than states that are closer to pure states. The pulse sequence for some states essentially performs a CNOT gate to create and remove entanglement; thus imperfect state preparation and measurement contributes significantly to the overall infidelity.
Using the data for 0, 1, and 2 gates, we can estimate the fidelity of a single CNOT gate normalized with respect to the overall fidelity of the state preparation and measurement steps. Assuming that the fidelity of the $n$th gate is $F_p^n = F_i (F_g)^n$, where $F_i$ is the preparation and measurement fidelity, the fitted fidelity per gate, $F_g$, is 95\%. 

\begin{table}[!htb]
\centering\begin{tabular}{| l | c | c |}
\hline
Gate & $F_p$ (\%) & $F_{\rm mean}$ (\%) \\ \hline
Identity & 90(1) & 94(3) \\          % "PT_ID.2", Jun 28, 09
CNOT   & 85(1) & 91(5) \\        % "PT_CNOT.2", Oct 25, 08 or "PT_CNOT", Jun 28, 09
CNOTx2 & 81(1) & 89(6) \\ \hline   % "PT_CNOTx2", Jun 28, 09
\end{tabular}
\caption{\label{Tbl:Data} Measured gate fidelities for the identity gate, the single CNOT gate, and two concatenated CNOT gates.}
\end{table}

\subsection{Error sources}  %%%%%%%%%%%%%%%%%%%%
\label{sec:gateerr}

A number of possible error sources and their contributions to the process fidelity of the single CNOT gate are listed in Table \ref{Tbl:errors}. To estimate and understand error sources, we simulated the full system evolution in the (2 atomic state) $\times$ (3 motional state) manifold using the exact Hamiltonian, including Stark shift and tomographic measurements. The magnitude of each source is measured independently and then added to the simulated pulse sequence. Laser frequency fluctuation is assumed to be the primary cause of decoherence and is measured by observing the decay of Ramsey fringes on the carrier transition. The frequency fluctuation is simulated as a random variable on the laser frequency which grows in amplitude over time, and accounted for via Monte Carlo techniques. Laser intensity fluctuations are measured directly with a photodiode. On short time scales comparable to the length of the gate, the fluctuations are $\sim 0.1\%$ peak to peak; on longer time scales, up to 1\% drifts are observed. Both of these effects are accounted for in the simulation. Off-resonant excitations are automatically included in the model of the full Hamiltonian. The effect can be removed from the simulation if decoherence is not included and the simulated pulses are of arbitrarily long lengths, which is equivalent to reducing the laser intensity. The resulting $\chi$ matrix and fidelity agree well with the measured results, indicating that the observed fidelity is well understood in terms of technical limitations. 

\begin{table}[!tb]
\centering\begin{tabular}{|l | c|c|}
\hline
Error source & Magnitude & Approx. contribution \\
\hline
Off-resonant excitations     & 1\% & 10\% \\
Laser frequency fluctuations & 300~Hz & 5\% \\
Laser intensity fluctuations & 1\% & 1\% \\ \hline
Total & & 15\% \\ \hline
\end{tabular}
\caption{\label{Tbl:errors} Error budget listing the major sources of errors on the process fidelity of the single CNOT gate, obtained by simulation. Each error source is assumed to be independent. The total error is calculated as the product of individual errors. }
\end{table}

Off-resonant excitations, caused by the square pulse shape used to address all transitions, is expected to be the largest source of error, as previous work has found \cite{Riebe:06}. Square pulses on the blue-sideband transition contain many higher harmonics, which causes residual excitation of the carrier transition. The carrier transition oscillations caused by this can be measured directly, averaged over many scans because of their small amplitude. Although the measured amount of off-resonant excitation is small ($\sim$1\%) for the laser intensity and secular frequencies used for our gates, both our simulations and previous work \cite{Riebe:06} have found that up to 10\% improvement in gate fidelity can be gained by implementing amplitude pulse shaping. 

\section{Conclusion}  %%%%%%%%%%%%%%%%%%%%%%%%%

In summary, we have developed a cryogenic, microfabricated ion trap system and demonstrated coherent control of a single ion. The cryogenic environment suppresses anomalous heating of the motional state, as well as enableing the use of a compact form of magnetic field stabilization using superconducting rings. We perform Stark shift correction in the pulse sequencer, removing the requirement for a separate laser path and acousto-optical modulator. A complete set of pulse sequences for performing quantum process tomography on a single ion's atomic and motional state is implemented. These components are sufficient to perform a CNOT gate on the atomic and motional state of a single ion. It is expected that amplitude pulse shaping would further improve the gate fidelity. These techniques, realized in a relatively simple experimental system, make the single ion a possible tool for studying other interesting quantum-mechanical systems.

The control techniques and the CNOT gate demonstrated in this work focus on a single ion and do not constitute a universal gate set for scalable quantum computation. However, the additional requirements for such a universal two-ion gate, including individual addressing \cite{Wang:09} and readout of two ions, have been realized in traditional 3D Paul traps as well as other surface trap experiments, and are not expected to pose significant challenges. The microfabricated surface-electrode ion trap operated in a cryogenic environment thus offers a viable option for realizing a large-scale quantum processor.

\section*{Acknowledgements}

We thank Eric Dauler for assistance in trap fabrication and Peter Herskind for helpful discussions and a critical reading of the manuscript. This work was supported by the Japan Science and Technology Agency, the COMMIT Program with funding from IARPA, and the NSF Center for Ultracold Atoms.

\bibliography{refs}

\end{document}